\documentclass[a4,article,nofootinbib]{revtex4}

\usepackage{graphicx}
\usepackage{dcolumn}
\usepackage{bm}
\usepackage{color}
\usepackage{amsmath}%
\usepackage{amsfonts}%
\usepackage{amssymb}%
\usepackage{breqn}
\usepackage{fixfoot}

\usepackage{bm}
\usepackage{mathrsfs}
\usepackage{amsthm}
\usepackage{breqn}
\usepackage{doi}
\usepackage{float}
\usepackage[normalem]{ulem}

\newcommand{\mc}{\mathcal}
\newcommand{\cp}{\times}

\newcommand{\bol}{\boldsymbol}

\newcommand{\abs}[1]{\left\lvert{#1}\right\rvert}
\newcommand{\w}{\wedge}
\newcommand{\lr}[1]{\left({#1}\right)}

\newcommand{\mf}{\mathfrak}
\newcommand{\p}{\partial}

\makeatletter
\let\cat@comma@active\@empty
\makeatother

\begin{document}
\title{Poissonization of Three Dimensional Nonholonomic Dynamics with the Method of Extension}
\author{N. Sato}
\affiliation{Research Institute for Mathematical Sciences, Kyoto University, Kyoto 606-8502, Japan}
\date{\today}

\begin{abstract}
In this study we develop a systematic procedure to construct 
a Poisson operator that describes the dynamics of a three dimensional nonholonomic system.  
Instead of reducing by symmetry the antisymmetric operator that links the energy gradient
to the velocity on the tangent bundle, the system is embedded in a larger space.
Here, the extended antisymmetric operator, which preserves the original equations of motion, satisfies
the Jacobi identity in a conformal fashion. Thus, a Poisson operator can be obtained by 
a further time reparametrization.  
Such `Poissonization' does not rely on the specific form of the Hamiltonian function.
The theory is applied to calculate the equilibrium distribution function of a
non-Hamiltonian ensemble.
\end{abstract}

\maketitle

\begin{normalsize}

\section{Introduction}
\label{intro}

Nonholonomic dynamical systems \cite{Bloch} can be characterized 
in terms of an energy integral (Hamiltonian function)  
and an antisymmetric operator (or almost Poisson operator) that fails to satisfy the 
Jacobi identity of Hamiltonian mechanics \cite{Mor}. 
These systems arise in the presence of non-integrable constraints \cite{Frankel} 
that limit the accessible regions of the phase space \cite{Schaft}.
In addition to the constrained motion of rolling rigid bodies \cite{deLeon},
and certain systems in molecular dynamics \cite{Sergi_1}, 
the Landau-Lifshitz equation of ferromagnetism \cite{Landau_2},
and the $\bol{E}\cp\bol{B}$ (E cross B) drift equation of plasma dynamics \cite{Cary}  
are examples in which the Jacobi identity can be violated \cite{Sato6}.

Such violation not only prevents the existence of phase space variables,
but has a critical role in determining the dynamical properties of the system \cite{Chandre}.  
In a recent paper \cite{Sato6} we have further shown that the rupture of the Hamiltonian structure
has fundamental implications for the formulation of statistical mechanics,
because one cannot rely on the invariant measure provided by Liouville's theorem in the Hamiltonian setting \cite{Sato}.

In order to formulate a statistical theory of conservative systems (i.e. systems that admit an energy integral but do not necessarily satisfy the Jacobi identity) it is therefore natural to explore the possibility of `repairing' the antisymmetric operator governing the
equations of motion, and thus recover phase space coordinates. 
In doing so, the method of reduction by symmetry \cite{Bates,Bloch2,GarciaN2007,Balseiro2} is not particularly helpful because it relies on the specific form of the Hamiltonian function. 

Here, we propose an alternative strategy that consists in embedding the system
in a larger space by introducing new degrees of freedom. The resulting extended antisymmetric operator 
is constructed so that the original equations of motion are preserved, while
the Jacobi identity is fulfilled by appropriately choosing a conformal factor \cite{Chaplygin,Balseiro}.
This factor (a strictly positive real valued function) then determines 
a time reparametrization by which the system
obeys Hamilton's canonical equations in an extended phase space.  
The procedure developed here applies to any three dimensional nonholonomic system.

The present paper is organized as follows. 
In section 2 we introduce the notation used throughout the manuscript.
In section 3 we discuss a three dimensional conservative system encountered in plasma dynamics, 
the so called $\bol{E}\cp\bol{B}$ drift equation of motion.
The procedure to Poissonize three dimensional conservative systems is developed in section 4,
and then applied in section 5 to Poissonize the $\bol{E}\cp\bol{B}$ drift dynamics of section 3.
Finally, in section 5 we obtain the equilibrium distribution function of an ensemble
of charged particles performing $\bol{E}\cp\bol{B}$ drift by using the canonical phase space recovered by Poissonization.

\section{Generalities}
\label{sec:1}

Let $\lr{x^{1},...,x^{n}}$ be a coordinate system on a smooth manifold $\mc{M}$ of dimension $n$
with tangent basis $\lr{\p_{1},...,\p_{n}}$.  
A conservative vector field $X\in T\mc{M}$ 
is defined as:
\begin{equation}
X=\mc{J}\lr{dH}=\mc{J}^{ij}H_{j}\p_{i},\label{eq1}
\end{equation}
Here the subscript indicates derivation. 
The bivector $\mc{J}\in\bigwedge^{2}T\mc{M}$ with $\mc{J}^{ij}=-\mc{J}^{ji}\in C^{\infty}\lr{\mc{M}}$ 
is called antisymmetric operator, and the real valued function $H\in C^{\infty}\lr{\mc{M}}$ the Hamiltonian. Due to antisymmetry, the Hamiltonian $H$ is a constant of motion, $\mf{L}_{X}H=0$. 
This is the reason why $X$ is conservative.
Nonholonomic mechanical systems admit the representation of equation \eqref{eq1} and are
characterized by the violation of the Jacobi identity, which demands the vanishing of the quantity
\begin{equation}
h=\mc{J}^{im}\mc{J}^{jk}_{m}+\mc{J}^{jm}\mc{J}^{ki}_{m}+\mc{J}^{km}\mc{J}^{ij}_{m}.\label{eq2}
\end{equation}
Hence, a nonholonomic system satisfies $h\neq 0$ somewhere on $\mc{M}$.
When $h=0$, the antisymmetric operator becomes a Poisson operator, and $X$ qualifies as an Hamiltonian vector field.  

In the case $n=3$ of $\mathbb{R}^{3}$, one can write:
\begin{equation}
\bol{v}=\bol{w}\cp\nabla H,\label{eq3}
\end{equation}
where $\bol{v}=\lr{\dot{x},\dot{y},\dot{z}}$, and $\bol{w}$ is a vector field such that $w_{x}=\mc{J}^{32}$, $w_{y}=\mc{J}^{13}$, and $w_{z}=\mc{J}^{21}$.
Now the Jacobi identity holds when the quantity:
\begin{equation}
h=\bol{w}\cdot\nabla\cp\bol{w},\label{eq4}
\end{equation}
vanishes. Notice that $\bol{w}\cdot\nabla\cp\bol{w}$ is nothing but the helicity density of $\bol{w}$.
In the following we shall refer to $h$ as the helicity density of $\mc{J}$.

System \eqref{eq3} is always subject to the constraint $\bol{w}\cdot\bol{v}=0$.
This constraint is integrable in the sense of the Frobenius' theorem \cite{Frankel} provided that $h=0$.
Hence, \eqref{eq3} is Hamiltonian if and only if the constraint is integrable.
Integrability implies that locally one has $\bol{w}=\lambda\nabla C$ for some functions
$\lambda$ and $C$. Then, $C$ is a constant of motion called Casimir invariant.
Furthermore, if $\bol{w}=\lambda\nabla C\neq \bol{0}$, the volume form $\lambda^{-1}dx\w dy\w dz$ is an invariant measure because $\nabla\cdot\lr{\lambda^{-1}\bol{v}}=0$. Conversely, if one can find a function $g\neq 0$ such that $\nabla\cdot\lr{g\bol{v}}=0$ for any choice of $H$, there exists a function $C$ such that $\bol{w}=g^{-1}\nabla C$, i.e. $\bol{w}$ is integrable. We conclude that, when $n=3$, the validity of the Jacobi identity,
the existence of a Casimir invariant, and the existence of an invariant measure for any choice of $H$, are
locally equivalent conditions.

\section{The Non-Hamiltonian Plasma Particle}
\label{sec:2}

Consider a charged particle moving in a static magnetic field $\boldsymbol{B}$ and under the influence of an electric field $\boldsymbol{E}=-\nabla\phi$. The equation of motion is:
\begin{equation}
m\dot{\bol{v}}=Z\left({\bol{v}}\cp\bol{B}+\bol{E}\right).\label{eq5}
\end{equation}    

\noindent Here $m$ is the particle mass and $Z$ the electric charge.
Suppose that $m$ is sufficiently small so that the left-hand side of equation (\ref{eq5}) can be neglected.
If we further take the cross product with the magnetic field $\boldsymbol{B}$, (\ref{eq5}) becomes:

\begin{equation}
\bol{v}_{\perp}=\frac{\boldsymbol{E}\times\boldsymbol{B}}{B^{2}},\label{eq6}
\end{equation}

\noindent where $\bol{v}_{\perp}$ is the velocity in the direction perpendicular to the magnetic field. We will also assume that the particle does not move along the magnetic
field, i.e. $\bol{v}_{\parallel}=\bol{v}-\bol{v}_{\perp}=\bol{0}$, giving $\bol{v}=\bol{v}_{\perp}$. The motion resulting from equation (\ref{eq6}) goes under the name of
$\bol{E}\cp\bol{B}$ drift and is the physical motivation behind the present study. 
We refer the reader to \cite{Cary} for a systematic derivation.

To simplify the notation we set $Z=1$. 
Recalling that the particle mass is small, the Hamiltonian of the system is $H=\phi$. 
The antisymmetric operator is then $\boldsymbol{w}=\boldsymbol{B}/B^{2}$, giving \eqref{eq6} in the form $\bol{v}=\bol{w}\cp\nabla H$. For this system to be Hamiltonian the Jacobi identity has to be satisfied. 
In light of \eqref{eq4}, this occurs only if (locally) $\boldsymbol{w}=\lambda\nabla C$, i.e. when the magnetic field is a solution to the equation $\bol{B}=B^{2}\lambda\nabla C$ for some appropriate $\lambda$ and $C$. 
The condition above is verified, for example, in the presence of an harmonic magnetic field $\boldsymbol{B}=\nabla\xi$. In this scenario $\lambda=B^{-2}$ and $C=\xi$. However, the Jacobi identity does not hold in the presence of a non-integrable magnetic field, implying that $\bol{E}\cp\bol{B}$ drift motion in a magnetic field with finite helicity $\bol{B}\cdot\nabla\cp\bol{B}\neq 0$ cannot be Hamiltonian. 

Below we give two examples that highlight the intrinsic difference between
the Hamiltonian and the non-Hamiltonian dynamical settings. 
First, consider the following system representing the motion of a plasma particle performing $\bol{E}\cp\bol{B}$ drift:
\begin{subequations}\label{eq7}
\begin{align}
&\bol{w}=\left(\cos z+\sin z\right)\p_{x}+\left(\cos z-\sin z\right)\p_{y},\label{thetaNhP}\\
&H=\frac{1}{2}\left(x^{2}+y^{2}+z^{2}\right).\label{H1}
\end{align}
\end{subequations} 
The equations of motion read:
\begin{dmath}
\bol{v}=\left(\cos z-\sin z\right)z\partial_{x}-\left(\cos z+\sin z\right)z\partial_{y}+\left[\left(\cos z+\sin z\right)y-\left(\cos z-\sin z\right)x\right]\partial_{z}.\label{eq8}
\end{dmath}
One can verify that $h=2\neq 0$, hence equation \eqref{eq8} is not Hamiltonian with non-integrable constraint $\bol{w}\cdot\bol{v}=0$.

Next, consider the motion of a rigid body 
with angular momentum $\boldsymbol{x}$ and momenta of inertia $I_{x},I_{y},I_{z}$: 
\begin{subequations}\label{eq9}
\begin{align}
&\bol{w}=x\p_{x}+y\p_{y}+z\p_{z},\\
&H=\frac{1}{2}\left(\frac{x^{2}}{I_{x}}+\frac{y^{2}}{I_{y}}+\frac{z^{2}}{I_{z}}\right).
\end{align}
\end{subequations}
This time the Jacobi identity is satisfied since $\nabla\cp\bol{w}=\bol{0}$, and the relevant Casimir invariant
is the total angular momentum $C=\boldsymbol{x}^{2}/2$, with $\bol{w}=\nabla C$. Thus, this second system is Hamiltonian, with integrale constraint $C$=constant. The equations of motion are:
\begin{equation}
\bol{v}=yz\left(\frac{1}{I_{z}}-\frac{1}{I_{y}}\right)\partial_{x}+xz\left(\frac{1}{I_{x}}-\frac{1}{I_{z}}\right)\partial_{y}+xy\left(\frac{1}{I_{y}}-\frac{1}{I_{x}}\right)\partial_{z}.\label{eq10}
\end{equation}

Figure \ref{fig1} shows the trajectory of the plasma particle (\ref{eq8}) and that of the rigid body (\ref{eq10}). Both of them lie on the integral surface of constant energy. However, while the orbit of the rigid body is closed and results from the intersection of the two integral manifolds defined by $H$ and $C=\boldsymbol{x}^{2}/2$, the plasma particle spirals toward a sink and delineates an open path characterized by the non-zero divergence of the conservative vector field (\ref{eq8}). 

\begin{figure}[h]
\hspace*{0cm}\centering
\includegraphics[scale=0.36]{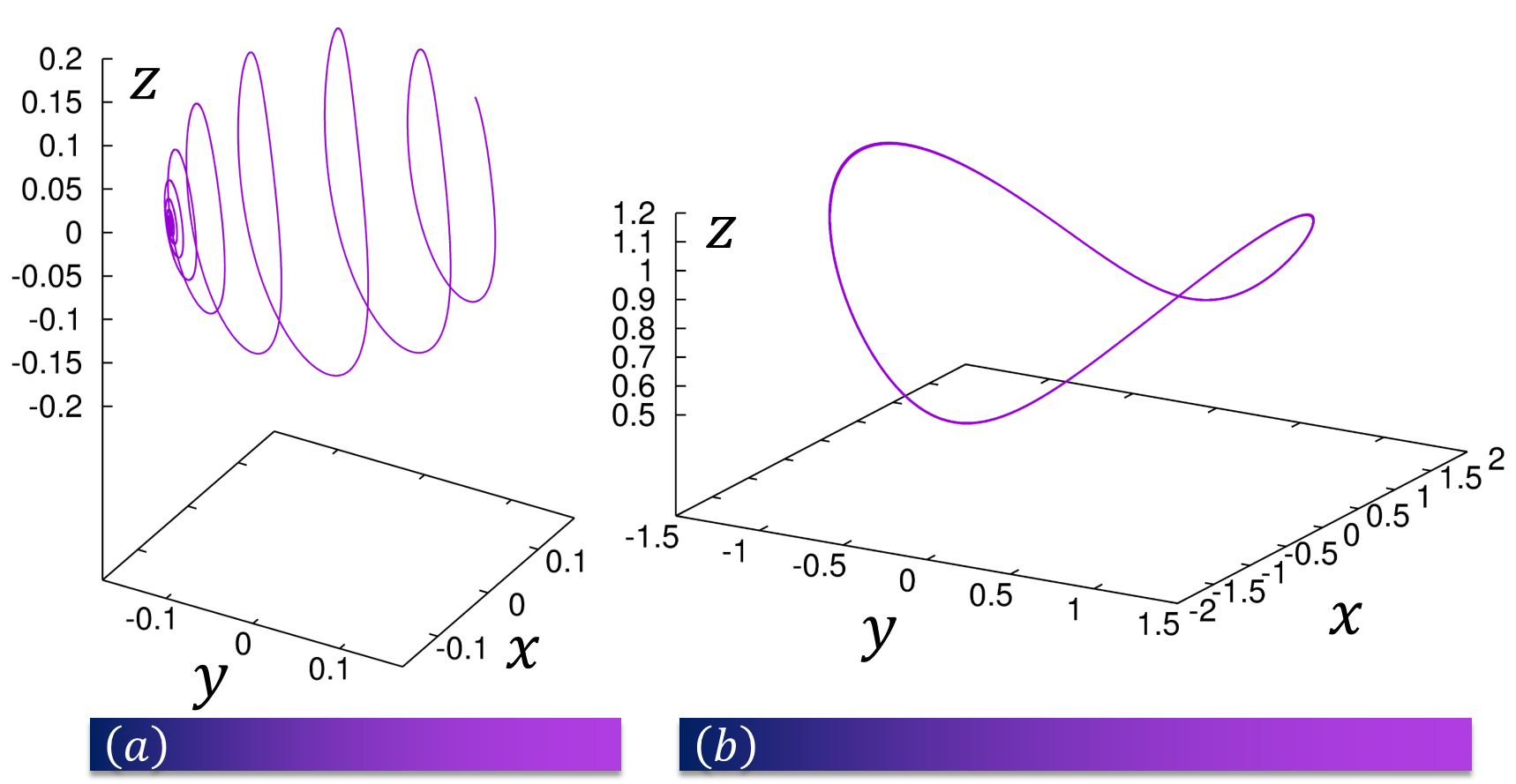}
\caption{\footnotesize (a): numerical integration of (\ref{eq8}). (b): numerical integration of (\ref{eq10}).}
\label{fig1}
\end{figure}	

As discussed at the end of section 2, this example shows that there is an important relationship between the
existence of an invariant measure and the Hamiltonian nature of the system. 
The absence of an invariant measure may be interpreted as the consequence of missing degrees of freedom that would compensate the compressibility of the system. This is why we will need to `extend' the system in order to recover an Hamiltonian structure.

\section{Poissonization in Three Dimensions}
\label{sec:3}

The purpose of the present section is to develop a systematic procedure to `repair' 
an arbitrary $3$-dimensional antisymmetric operator and obtain an equivalent Hamiltonian system
describing the same dynamics.

\subsection{\textit{Extension}}

The first step of the procedure consists in the embedding of the system in a larger space.
The objective is to restore 
a conformally Poisson structure. 
To do so in the three dimensional setting, it will be sufficient to add a single new variable $s$.
We begin by extending the antisymmetric operator $\mathcal{J}$ in the following manner:
\begin{equation}
\mathfrak{J}=\mathcal{J}+a\,\partial_{x}\wedge\partial_{s}+b\,\partial_{y}\wedge\partial_{s}+c\,\partial_{z}\wedge\partial_{s},\label{eq11}
\end{equation} 
where $\mathfrak{J}$ is the $4\times 4$ extended antisymmetric operator and the coefficients $a,b,c$ have to be determined by requiring that the new operator is conformally Poisson. 
We remark that these new terms do not affect the
original equations of motion since the Hamiltonian function does not depend on the new variable $s$, i.e. $H_{s}=0$.

An antisymmetric operator $\mf{J}$ is a conformally Poisson operator if there exists a strictly
positive real valued function $r$ such that $r^{-1}\mf{J}$ satisfies the Jacobi identity.
If the matrix $\mf{J}^{ij}$ is invertible with inverse $\Omega_{ij}$ such that $i_{X}\Omega=-dH$ with $\Omega=\sum_{i<j}\Omega_{ij}dx^{i}\wedge dx^{j}\in \bigwedge^{2}T^{\ast}\mc{M}$, this condition is equivalent to demanding that:
\begin{equation}
d\left(r\Omega\right)=0,\label{eq12}
\end{equation}
for some conformal factor $r> 0$. 
%
%
Looking for the inverse of (\ref{eq11}), we find:
\begin{dmath}\label{eq13}
\Omega=\frac{1}{aw_{x}+bw_{y}+cw_{z}}\left\{
\left[a-s\left(\frac{\partial w_{z}}{\partial y}-\frac{\partial w_{y}}{\partial z}\right)\right]dy\wedge dz
+\left[b-s\left(\frac{\partial w_{x}}{\partial z}-\frac{\partial w_{z}}{\partial x}\right)\right]dz\wedge dx
+\left[c-s\left(\frac{\partial w_{y}}{\partial x}-\frac{\partial w_{x}}{\partial y}\right)\right]dx\wedge dy
+d\left(w_{x}s\right)\wedge dx+d\left(w_{y}s\right)\wedge dy+d\left(w_{z}s\right)\wedge dz\right\}.
\end{dmath}

\noindent With the choice:

\begin{subequations}\label{eq14}
\begin{align}
&a=D_{x}+s\left(\frac{\partial w_{z}}{\partial y}-\frac{\partial w_{y}}{\partial z}\right),\\
&b=D_{y}+s\left(\frac{\partial w_{x}}{\partial z}-\frac{\partial w_{z}}{\partial x}\right),\\
&c=D_{z}+s\left(\frac{\partial w_{y}}{\partial x}-\frac{\partial w_{x}}{\partial y}\right),
\end{align}
\end{subequations}

\noindent the $2$-form $\Omega$ becomes:

\begin{equation}\label{eq15}
\Omega=\frac{\mathcal{D}+d\left[s\lr{w_{x}dx+w_{y}dy+w_{z}dz}\right]}
{\bol{w}\cdot\lr{\bol{D}+s\nabla\cp\bol{w}}}
=\frac{d\lr{\mc{A}+s\theta}}{\ast\left[\theta\wedge\lr{\mathcal{D}+sd\theta}\right]},
\end{equation}

\noindent Here $\theta=w_{x}dx+w_{y}dy+w_{z}dz$, $\mathcal{D}=d\mathcal{A}=D_{x}dy\wedge dz+D_{y}dz\wedge dx+D_{z}dx\wedge dy$ is an arbitrary closed $2$-form that does not depend on $s$, and $\boldsymbol{D}=\left(D_{x},D_{y},D_{z}\right)$. 
From equation (\ref{eq15}) we see that $\Omega$ is conformally closed with conformal factor:

\begin{equation}
r=
\abs{\bol{w}\cdot\bol{D}+sh}=\abs{\ast \lr{\theta\wedge\mc{D}}+sh}
,\label{eq16}
\end{equation}
as long as $r\neq 0$. We will see how to physically determine $\bol{D}$ in the following section.
At this point just notice that by appropriately choosing $\bol{D}$ one can always set $r>0$.


The extended equations of motion $X=\mathfrak{J}\left(dH\right)$ now read:

\begin{equation}\label{eq17}
X=\boldsymbol{w}\times\nabla H-\left(\boldsymbol{D}+s\nabla\times\boldsymbol{w}\right)\cdot\nabla H\,\partial_{s}.
\end{equation}

\noindent Finally, one can verify that the new equations are divergence free (the extended antisymmetric operator guarantees an invariant measure for any choice of $H$):

\begin{equation}\label{eq18}
{div}\left(X\right)=\nabla\cdot\left(\boldsymbol{w}\times\nabla H\right)-\nabla\times\boldsymbol{w}\cdot\nabla H=0.
\end{equation}

\subsection{\textit{Time reparametrization}}

The second step of the procedure involves a time reparametrization that will give us the desired Poisson structure. This result can be achieved by introducing the new time variable (proper time) $\tau$ satisfying:

\begin{equation}\label{eq19}
\frac{d\tau}{dt}=r.
\end{equation}

\noindent Since by construction $r^{-1}\mf{J}$ satisfies the Jacobi identity, the vector field:
\begin{equation}\label{eq20}
Y=r^{-1}X=r^{-1}\mf{J}\lr{dH}=r^{-1}\frac{dx^{i}}{dt}\p_{i}=\frac{dx^{i}}{d\tau}\p_{i},
\end{equation}
is an Hamiltonian vector field expressing the equations of motion with respect to the proper time $\tau$.

\section{Poissonization of the Non-Hamiltonian Plasma Particle}
\label{sec:4}

\noindent Here we apply the developed procedure to Poissonize the $\bol{E}\cp\bol{B}$ drift motion of the plasma particle
studied in section $3$. 

\subsection{\textit{The physical meaning of $s$ and $\tau$}}

First, let us spend some words on the physical meaning of the new variable $s$.
From equations \eqref{eq7} and (\ref{eq17}), and noting that in this case $\boldsymbol{w}=\boldsymbol{B}/B^{2}$ with $B^{2}=1/2$, we have:

\begin{equation}
\begin{split}
\dot{s}&=-\left(\boldsymbol{D}+s\nabla\times\boldsymbol{w}\right)\cdot\nabla H=-\left(\boldsymbol{D}+s\boldsymbol{w}\right)\cdot\nabla H\\&=-\left(\boldsymbol{B}+s\frac{\boldsymbol{B}}{B^{2}}\right)\cdot\nabla H=-\left(\frac{1}{\sqrt{2}}+s\sqrt{2}\right)\frac{\partial H}{\partial \ell},\label{eq21}
\end{split}
\end{equation}

\noindent Here we used the fact that $\bol{w}=\nabla\cp\bol{w}$ and made the choice $\boldsymbol{D}=\boldsymbol{B}$ (we will justify this choice later). The variable $\ell$ measures the length along a field line ($\partial_{\ell}=\boldsymbol{B}/B$). 
%
%
We define:

\begin{equation}
m\tilde{v}_{\parallel}=\frac{1}{\sqrt{2}}\log{\left[c_{0}\left(\frac{1}{\sqrt{2}}+s\sqrt{2}\right)\right]},\label{eq22}
\end{equation}

\noindent with $c_{0}$ a constant. This implies:

\begin{equation}
m\tilde{v}_{\parallel}=-\frac{\partial H}{\partial \ell}.\label{eq23}
\end{equation} 

\noindent Thus, the new variable $\tilde{v}=d\ell/dt$ has the simple interpretation of a velocity in the direction parallel to $\boldsymbol{B}$: the missing degree of freedom $s$ just describes the motion along the magnetic field which was neglected in the original three dimensional description of the dynamics. 
Inverting equation (\ref{eq22}) we also have:

\begin{equation}\label{eq24}
s=\frac{1}{2}\left(e^{\sqrt{2}m\tilde{v}_{\parallel}}-1\right).
\end{equation}

\noindent In the above equation we required that $s=0$ when $\tilde{v}_{\parallel}=0$ so that $c_{0}=\sqrt{2}$ (we will justify this choice later).  

What about the meaning of the proper time $\tau$? Using the expression for $r$ equation (\ref{eq16}),

\begin{equation}
r=1+sh=1+2s,\label{eq25}
\end{equation}

\noindent Here we used the fact that $\boldsymbol{D}\cdot\boldsymbol{w}=\boldsymbol{B}\cdot\boldsymbol{w}=1$. 
From (\ref{eq19}):

\begin{equation}\label{eq26}
\frac{d\tau}{dt}=1+2s=e^{\sqrt{2}m\tilde{v}_{\parallel}}
\end{equation}

\noindent Thus, the choices $\boldsymbol{D}=\boldsymbol{B}$ and $c_{0}=\sqrt{2}$ are now physically justified because the factor $r$ must be $1$ when $\boldsymbol{w}$ is integrable or $\tilde{v}_{\parallel}=0$, i.e. we must have $d\tau/dt=1$ when the Jacobi identity is satisfied or there is no motion along the magnetic field.  
If the mass $m$ is small we can expand the exponential to obtain:

\begin{equation}\label{eq27}
\frac{d\tau}{dt}=1+\sqrt{2}m\tilde{v}_{\parallel}+o\left(\left(\sqrt{2}m\tilde{v}_{\parallel}\right)^{2}\right).
\end{equation}

\noindent Neglecting second order terms and noting that $\tilde{v}_{\parallel}\simeq d\ell/dt$, the result is:

\begin{equation}\label{eq28}
\tau\simeq t+\sqrt{2}m\ell,
\end{equation}

\noindent and the proper time $\tau$ can be interpreted as a measure of the distance traveled by the particle along the magnetic field. Note however that the dynamics described by $s$ does not affect 
the `real' orbit of the particle in $\mathbb{R}^{3}$. Therefore, the distance $\ell$ is `fictitious'. 

\subsection{\textit{Poissonization in Cartesian coordinates}}

We are now ready to write the canonical equations of motion for the plasma particle.
Recalling (\ref{eq15}), the symplectic $2$-form of interest is:

\begin{dmath}\label{eq29}
\Omega^{'}=r\Omega=d\left[\mathcal{A}+s\left(\cos z+\sin z\right)dx+s\left(\cos z-\sin z\right)dy\right]=
-d\left\{xd\left[\left(s+\frac{1}{2}\right)\left(\cos z+\sin z\right)\right]+yd\left[\left(s+\frac{1}{2}\right)\left(\cos z-\sin z\right)\right]\right\},
\end{dmath}

\noindent where we used the fact that $\mathcal{A}=\theta/2$. Thus, the canonical variables are:

\begin{subequations}\label{eq30}
\begin{align}
&q_{x}=\left(s+\frac{1}{2}\right)\left(\cos z+\sin z\right),\label{qx}\\
&p_{x}=-x,\label{px}\\
&q_{y}=\left(s+\frac{1}{2}\right)\left(\cos z-\sin z\right),\label{qy}\\
&p_{y}=-y.\label{py}
\end{align}
\end{subequations} 

\noindent In terms of these new variables we also have:

\begin{subequations}\label{eq31}
\begin{align}
&z=\arcsin\left[\frac{q_{x}-q_{y}}{\sqrt{2\left(q^{2}_{x}+q^{2}_{y}\right)}}\right],\\
&s+\frac{1}{2}=\sqrt{\frac{q^{2}_{x}+q^{2}_{y}}{2}},\\
&H=\frac{1}{2}\left(p^{2}_{x}+p^{2}_{y}+\arcsin^{2}\left[\frac{q_{x}-q_{y}}{\sqrt{2\left(q^{2}_{x}+q^{2}_{y}\right)}}\right]\right).
\end{align}
\end{subequations}

\noindent Here, we chose the positive root for $s+1/2$. Finally, denoting with $'$ derivations with
repsect to $\tau$, Hamilton's canonical equations read:

\begin{subequations}\label{eq32}
\begin{align}
&q^{'}_{x}=H_{p_{x}}=p_{x},\\
&p^{'}_{x}=-H_{q_{x}}=\frac{-q_{y}}{q^{2}_{x}+q^{2}_{y}}\arcsin\left[\frac{q_{x}-q_{y}}{\sqrt{2\left(q^{2}_{x}+q^{2}_{y}\right)}}\right],\\
&q^{'}_{y}=H_{p_{y}}=p_{y},\\
&p^{'}_{y}=-H_{q_{y}}=\frac{q_{x}}{q^{2}_{x}+q^{2}_{y}}\arcsin\left[\frac{q_{x}-q_{y}}{\sqrt{2\left(q^{2}_{x}+q^{2}_{y}\right)}}\right].
\end{align}
\end{subequations}

\noindent Figure \ref{fig2} shows a numerical integration of the Hamiltonian system (\ref{eq32}). 
The solution progressively approaches a $2$-dimensional uniform rectilinear motion. 

\begin{figure}[h]
\hspace*{-0.1cm}\centering
\includegraphics[scale=0.26]{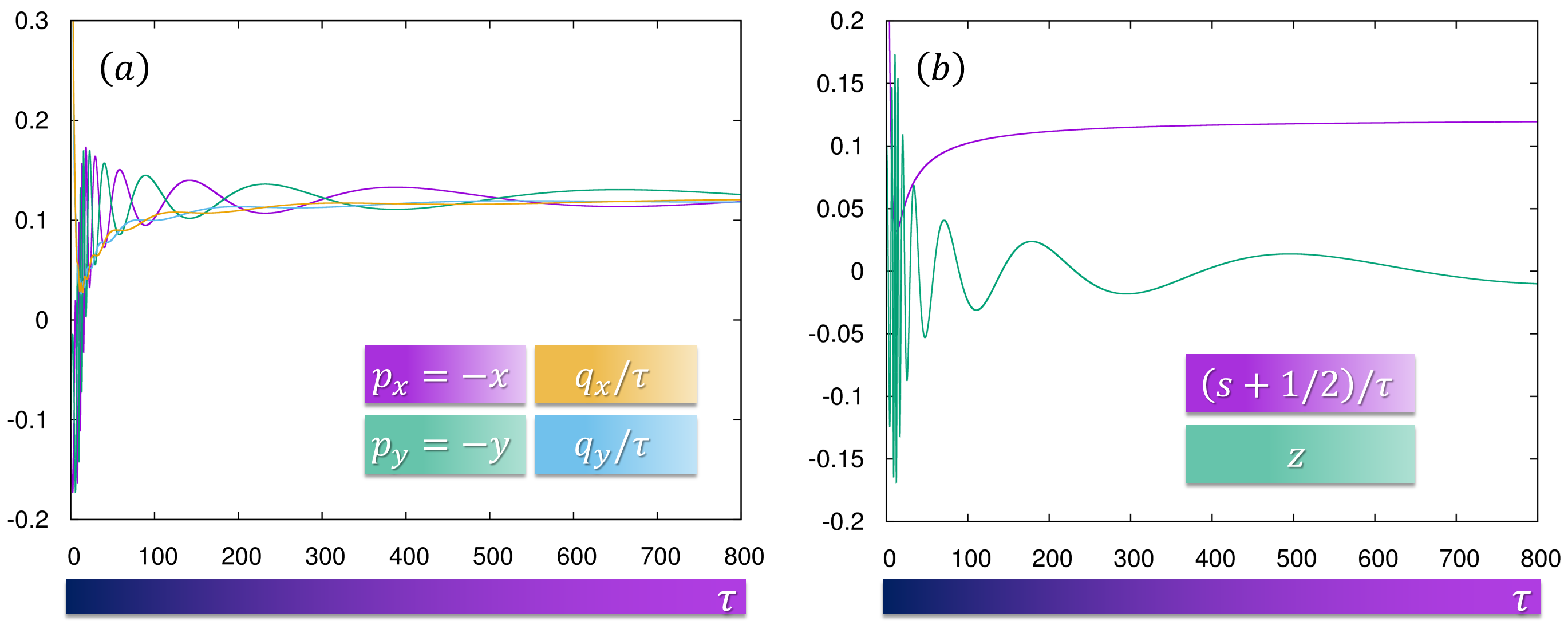}
\caption{\footnotesize Numerical integration of system (\ref{eq32}). (a): evolution with respect to the proper time $\tau$ of $p_{x}$, $q_{x}/\tau$, $p_{y}$, and $q_{y}/\tau$. (b): evolution with respect to the proper time $\tau$ of $\left(s+\frac{1}{2}\right)/\tau$ and $z$.}
\label{fig2}
\end{figure}	

Finally, observe that in the original time $t$, the equations of motion for the canonical variables $p_{x}$, $q_{x}$, $p_{y}$, and $q_{y}$ take the form:

\begin{subequations}\label{eq33}
\begin{align}
&\dot{q}_{x}=r^{-1}H_{p_{x}},\\
&\dot{p}_{x}=-r^{-1}H_{q_{x}},\\
&\dot{q}_{y}=r^{-1}H_{p_{y}},\\
&\dot{p}_{y}=-r^{-1}H_{q_{y}}.
\end{align}
\end{subequations}

\noindent These equations, which are not canonical, imply that the `force' acting on the particle is only proportional to the gradient of the Hamiltonian with proportionality factor $r^{-1}$. Therefore, the same energy gradient produces different forces depending on the position in space. Such behavior departs from the standard laws of physics and signals the importance of the Jacobi identity in determining the structure of the equations of motion. This inhomogeneity is also the reason why canonical equations can be obtained only by `adjusting' the time variable. 

\subsection{\textit{Poissonization in magnetic coordinates}}

Consider again the plasma particle of section $3$.
By performing an appropriate change of coordinates before the Poissonization procedure, we can simplify
the antisymmetric operator $\mathcal{J}$. 
The simplified form will offer us an insight into the relation between $\mathcal{J}$, which is the result of the
abrupt reduction $m\rightarrow 0$ together with $\bol{v}_{\parallel}\rightarrow\bol{0}$, and the full dynamics of a magnetized particle.   
The target coordinates are the so called magnetic coordinates $\left(\ell,\psi,\zeta\right)$ which we will define shortly. 
This time we shall consider a more general type of magnetic field\footnote{Note that with the identification $\ell=y$, $\psi=\left(\sin z+\cos z\right)/2$, $\zeta=x$, $i=\left(\sin z+\cos z\right)/\left(\sin z-\cos z\right)$, where $\left(x,y,z\right)$ are Cartesian coordinates, equation (\ref{eq34}) gives the magnetic field studied in the example of equation (\ref{eq7}).}:

\begin{equation}\label{eq34}
\boldsymbol{B}=\nabla\psi\times\nabla\zeta+i\left(\ell,\psi\right)\nabla\psi\times\nabla\ell.
\end{equation}

\noindent Here $i$ is an arbitrary function of $\ell$ and $\psi$. When $i=0$ equation (\ref{eq34}) can conveniently represent a dipole magnetic field. The flux function $\psi$ is chosen so that
$\psi=\psi\left(R,z\right)$, where $\left(R,\zeta,z\right)$ is a cylindrical coordinate system 
$dx\w dy\w dz\w= R \,dR\w d\zeta\w dz$ with $R$ the radial coordinate in the $\left(x,y\right)$ plane and $\zeta$ the toroidal angle. The coordinate $\ell$ is defined to be the length along the field lines of the poloidal component $\boldsymbol{B}_{p}=\nabla\psi\times\nabla\zeta$ of the magnetic field $\boldsymbol{B}$. In formulae:

\begin{equation}\label{eq35}
\partial_{\ell}=\frac{\boldsymbol{B}_{p}}{B_{p}}.
\end{equation}

\noindent Thus, if $i\neq 0$, the magnetic field has a toroidal component $\boldsymbol{B}_{t}=i\left(\ell,\psi\right)\nabla\psi\times\nabla\ell$. In terms of the new coordinates, the magnetic field $2$-form $\mathfrak{B}$ reads as:

\begin{equation}\label{eq36}
\mathfrak{B}=d\psi\wedge d\zeta+id\psi\wedge d\ell.
\end{equation} 

\noindent Note that $d\mathfrak{B}=0$. In order to express $\mathcal{J}$ in the new variables we need some geometrical relationships among tangent and cotangent vectors. We begin by calculating the Jacobian $\mathcal{Q}$ of the coordinate change: 

\begin{equation}\label{eq37}
\begin{split}
\mathcal{Q}&=\nabla\ell\cdot\nabla\psi\times\nabla\zeta=B_{p}.
\end{split}
\end{equation}

\noindent Here we used $\nabla\ell\cdot\partial_{\ell}=1$. Similarly, by using the reciprocity relationships
$\nabla u^j\cdot\partial_{u^k}=\delta_{jk}$ with $\lr{u^1,u^2,u^3}=\lr{\ell,\psi,\zeta}$, one obtains the following expressions:

\begin{subequations}\label{eq38}
\begin{align}
\nabla\ell&=\frac{1}{\left\lvert\partial_{\psi}\right\rvert^{2}-q^{2}}\left(q\partial_{\psi}+\left\lvert\partial_{\psi}\right\rvert^{2}\partial_{l}\right),\\
\nabla\psi&=\frac{1}{\left\lvert\partial_{\psi}\right\rvert^{2}-q^{2}}\left(\partial_{\psi}+q\partial_{l}\right),\\
\left\lvert\nabla\ell\right\rvert^{2}&=\frac{\left\lvert\partial_{\psi}\right\rvert^{2}}{\left\lvert\partial_{\psi}\right\rvert^{2}-q^{2}},\\
\left\lvert\partial_{\psi}\right\rvert^{2}&=\frac{\left\lvert\nabla\ell\right\rvert^{2}}{R^{2}B_{p}^{2}},\\
B^{2}_{p}&=\frac{1}{R^{2}\left(\left\lvert\partial_{\psi}\right\rvert^{2}-q^{2}\right)}.
\end{align} 
\end{subequations}

\noindent Here $q=-\partial_{\ell}\cdot\partial_{\psi}$. Recalling that the $\bol{E}\cp\bol{B}$ equation of
motion is given by (\ref{eq6}) and exploiting (\ref{eq38}) with the identities $\nabla u^j=\left(\partial_{u^k}\times\partial_{u^m}\right)/\left(\partial_{u^j}\cdot\partial_{u^k}\times\partial_{u^m}\right)$, $\partial_{u^j}=\left(\nabla{u^k}\times\nabla{u^m}\right)/\left(\nabla{u^j}\cdot\nabla{u^k}\times\nabla{u^m}\right)$ where $j,k,m$ are all different:

\begin{subequations}\label{eq39}
\begin{align}
\dot{\ell}&=\nabla\ell\cdot X=\rho\left(iR^{2}H_{\psi}-qH_{\zeta}\right),\\
\dot{\psi}&=\nabla\psi\cdot X=-\rho\left(H_{\zeta}+iR^{2}H_{\ell}\right),\\
\dot{\zeta}&=\nabla\zeta\cdot X=\rho\left(H_{\psi}+qH_{\ell}\right),
\end{align}
\end{subequations} 

\noindent where $\rho=B^{2}_{p}/B^{2}$. Therefore, the bivector $\mathcal{J}$ takes the form:

\begin{equation}\label{eq40}
\mathcal{J}=\rho\left(\partial_{\zeta}\wedge\partial_{\psi}+q\partial_{\zeta}\wedge\partial_{\ell}+iR^{2}\partial_{\ell}\wedge\partial_{\psi}\right).
\end{equation}

\noindent A straightforward application of the Poissonization procedure of section $4$ gives the
$4$-dimensional conformal $2$-form:

\begin{equation}\label{eq41}
\Omega=\frac{\mathfrak{B}+d\left(s\theta\right)}{\ast\left[\theta\wedge\left(\mathfrak{B}+sd\theta\right)\right]},
\end{equation}

\noindent where we set $\mathcal{D}=d\mathcal{A}=\mathfrak{B}$ and the kernel of $\mathcal{J}$ is spanned by the covector:

\begin{equation}\label{eq42}
\theta=\rho \left(d\ell-q d\psi-iR^{2} d\zeta\right).
\end{equation}   

\noindent Thus, a time reparametrization $d\tau/dt=r$ with conformal factor:

\begin{dmath}\label{eq43}
r=\ast\left[\theta\wedge\left(\mathfrak{B}+sd\theta\right)\right]=
\rho\left\{1+i^{2}R^{2}+s\left[\rho\frac{\partial q}{\partial\zeta}+iR^{2}\left(\frac{\partial\left(q\rho\right)}{\partial\ell}+\frac{\partial\rho}{\partial\psi}\right)-\left(q\partial_{\ell}+\partial_{\psi}\right)\left(iR^{2}\rho\right)\right]\right\},
\end{dmath}

\noindent will give the symplectic $2$-form:

\begin{dmath}\label{eq44}
\Omega^{'}=r\Omega=\mathfrak{B}+d\left(s\theta\right)=d\psi\wedge d\zeta+id\psi\wedge d\ell+d\left[s\rho\left( d\ell-q d\psi-iR^{2} d\zeta\right)\right].
\end{dmath}

\noindent Now suppose that $\boldsymbol{B}=\boldsymbol{B}_{p}$, i.e. $i=0$. Then, $\rho=1$ (note that $q_{\zeta}=0$ when $i=0$ due to toroidal symmetry), $\theta\wedge d\theta=(d\ell-qd\psi)\wedge\left(q_{\ell}d\ell\wedge d\psi\right)=0$ and $r=1$. Furthermore, $\Omega$ becomes symplectic:

\begin{equation}\label{eq45}
\Omega=d\left(\psi d\zeta+ sd\ell-sqd\psi\right).
\end{equation}  

\noindent With the identification $s=v_{\parallel}$ of equation (\ref{eq24}) for $m<<1$, the expression (\ref{eq45}) is exactly the symplectic $2$-form for the motion of a magnetized particle in a magnetic field of the form $\nabla\psi\times\nabla\zeta$ \cite{SatoSelf}. From this example we can see 
that the Poissonization procedure reproduces the correct physics, and that in the presence of a general magnetic field (note that (\ref{eq41}) holds for any $\mathfrak{B}$) canonical coordinates can be obtained by operating a time reparametrization. 

We conclude this section by giving the antisymmetric operator and the conformal factor for a magnetic field
written as:

\begin{equation}\label{eq46}
\boldsymbol{B}=\alpha\nabla\psi\times\nabla\zeta+i\nabla\psi\times\nabla\ell+\beta\nabla\zeta\times\nabla\ell.
\end{equation} 

\noindent Here $\alpha$, $\beta$, and $i$ are $3$ arbitrary functions satisfying $\alpha_{\ell}-i_{\zeta}+\beta_{\psi}=0$ to ensure that $d\mathfrak{B}=0$. 
With the same procedure as above one obtains:

\begin{equation}\label{eq47}
\mathcal{J}=\rho\left[\left(\alpha-\beta q\right)\partial_{\zeta}\wedge\partial_{\psi}+\left(\beta\left\lvert\partial_{\psi}\right\rvert^{2}-\alpha q\right)\partial_{\ell}\wedge\partial_{\zeta}+iR^{2}\partial_{\ell}\wedge\partial_{\psi}\right],
\end{equation}

\noindent The kernel of this operator is spanned by the covector:

\begin{equation}\label{eq48}
\theta=\rho\left[\left(\alpha-\beta q\right)d\ell+\left(\beta\left\lvert\partial_{\psi}\right\rvert^{2}-\alpha q\right)d\psi-iR^{2}d\zeta\right]
\end{equation}

\noindent The conformal factor is:

\begin{dmath}\label{eq49}
r=\rho\left\{\alpha\left(\alpha-\beta q\right)+\beta\left(\beta\left\lvert\partial_{\psi}\right\rvert^{2}-\alpha q\right)+i^{2}R^{2}+s\left[\left(\alpha-\beta q\right)\left(-\frac{\partial\left(iR^{2}\rho\right)}{\partial\psi}-\frac{\partial\rho\left(\beta\left\lvert\partial_{\psi}\right\rvert^{2}-\alpha q\right)}{\partial\zeta}\right)
+\left(\beta\left\lvert\partial_{\psi}\right\rvert^{2}-\alpha q\right)\left(\frac{\partial\rho\left(\alpha-\beta q\right)}{\partial\zeta}+\frac{\partial\left(iR^{2}\rho\right)}{\partial\ell}\right)
-iR^{2}\left(\frac{\partial\rho\left(\beta\left\lvert\partial_{\psi}\right\rvert^{2}-\alpha q\right)}{\partial\ell}-\frac{\partial\rho\left(\alpha-\beta q\right)}{\partial\psi}\right)
\right]\right\}.
\end{dmath}

\noindent Finally, the symplectic $2$-form recovered after time reparametrization is:

\begin{dmath}\label{eq50}
\Omega^{'}=\alpha d\psi\wedge d\zeta+id\psi\wedge d\ell +\beta d\zeta\wedge d\ell+d\left[s\rho\left(\left(\alpha-\beta q\right)d\ell+\left(\beta\left\lvert\partial_{\psi}\right\rvert^{2}-\alpha q\right)d\psi-iR^{2}d\zeta\right)\right].
\end{dmath}

\section{Statistical Mechanics of the Non-Hamiltonian Plasma Particle}
\label{sec:5}

In this section we apply the theory to the study of the statistical behavior of an ensemble of particles obeying equation \eqref{eq6}.

\subsection{\textit{The Jacobian of the coordinate change}}

By extending system \eqref{eq3} to $4$-dimensions $\boldsymbol{x}=\left(x,y,z,s\right)$, 
according to equation \eqref{eq18} one obtains the invariant measure (in time $t$):

\begin{equation}\label{eq51}
vol^{4}_{\boldsymbol{x}}=dx\w dy\w dz\w ds.
\end{equation}
 
\noindent A further time reparametrization, equation (\ref{eq19}), 
gives a symplectic manifold $\boldsymbol{y}=\left(p_{x},q_{x},p_{y},q_{y}\right)$:

\begin{equation}
vol^{4}_{\boldsymbol{y}}=dp_{x}\w dq_{x}\w dp_{y}\w dq_{y}.\label{eq52}
\end{equation}

\noindent The canonical variables $\boldsymbol{y}$ are determined by the specific form of the antisymmetric operator $\boldsymbol{w}$ so that $r\Omega=dp_{x}\wedge dq_{x}+dp_{y}\wedge dq_{y}$. 
We want to determine the Jacobian of the coordinate change sending (\ref{eq51}) to (\ref{eq52}).
For this purpose, we need the following:\\

\textit{Let $X=d\boldsymbol{x}/dt$ and $Y=d\boldsymbol{y}/d\tau$ be two vector fields with $\boldsymbol{x}=\left(x^{1},...,x^{n}\right)$ and $\boldsymbol{y}=\left(y^{1},...,y^{n}\right)$. Let $g$ be the Jacobian of the coordinate change $vol_{\boldsymbol{y}}^{n}=dy^{1}\wedge ... \wedge dy^{n}=g^{-1}dx^{1}\wedge ... \wedge dx^{n}=g^{-1}vol_{\boldsymbol{x}}^{n}$. If}
\begin{equation}\label{eq53}
\mathfrak{L}_{X}vol_{\boldsymbol{x}}^{n}=\mathfrak{L}_{Y}vol_{\boldsymbol{y}}^{n}=0,
\end{equation} 
\noindent \textit{then,}
\begin{equation}\label{eq54}
g=\frac{dt}{d\tau}.
\end{equation}

\noindent We have:
\begin{dmath}\label{eq55}
0=\mathfrak{L}_{X}vol_{\boldsymbol{x}}^{n}=
di_{X}gvol_{\boldsymbol{y}}^{n}=
(-1)^{m-1}d(g\dot{y}^{m})\wedge dy^{1}\wedge ... \wedge dy^{m-1} \wedge dy^{m+1}\wedge ... \wedge dy^{n}=
\frac{1}{g}\frac{\partial}{\partial y^{m}}\left(g\frac{d\tau}{dt}\left(y^{m}\right)'\right)vol_{\boldsymbol{x}}^{n}=\frac{d\tau}{dt}\frac{\partial\left(y^{m}\right)'}{\partial y^{m}}vol_{\boldsymbol{x}}^{n}+\frac{\left(y^{m}\right)'}{g}\frac{\partial}{\partial y^{m}}\left(g\frac{d\tau}{dt}\right)vol_{\boldsymbol{x}}^{n}=\frac{\left(y^{m}\right)'}{g}\frac{\partial}{\partial y^{m}}\left(g\frac{d\tau}{dt}\right)vol_{\boldsymbol{x}}^{n}=\frac{1}{g}\frac{d}{d\tau}\left(g\frac{d\tau}{dt}\right)vol_{\boldsymbol{x}}^{n}.
\end{dmath}
Here, the apex $'$ indicates derivation with respect to $\tau$ and we used the fact that $\mathfrak{L}_{Y}vol_{\boldsymbol{y}}^{n}=0$ if and only if $\partial_{y^{m}}\left(y^{m}\right)'=0$. 
The solution is, up to a scaling constant, $g=dt/ d\tau$. 

Applying this result to the case $\boldsymbol{x}=\left(x,y,z,s\right)$ and $\boldsymbol{y}=\left(p_{x},q_{x},p_{y},q_{y}\right)$ we conclude that the Jacobian $g$ of the coordinate change is:

\begin{equation}\label{eq56}
g=\frac{dt}{d\tau}=r^{-1}=\frac{1}{\boldsymbol{w}\cdot\bol{D}+sh}.
\end{equation}

\subsection{\textit{The distribution function of thermodynamic equilibrium}}

Let $P=P(\boldsymbol{y})$ be the distribution function of an ensemble of particles in the canonical phase space $vol_{\boldsymbol{y}}^{4}$ at thermodynamic equilibrium $\tau\rightarrow\infty$. 
We want to know how the distribution function $P$ is seen in the initial coordinates $vol_{\boldsymbol{x}}^{4}$.
Using the result of equation (\ref{eq56}), we have:

\begin{equation}\label{eq57}
P\,vol_{\boldsymbol{y}}^{4}=Pr\,vol_{\boldsymbol{x}}^{4},
\end{equation}

\noindent which implies that the distribution function $f\left(\boldsymbol{x}\right)$ on $vol_{\boldsymbol{x}}^{4}$ is related to $P$ as:

\begin{equation}\label{eq58}
f=Pr=P\lr{\bol{w}\cdot\boldsymbol{D}+sh}.
\end{equation}

\noindent From the result above we see that the discrepancy between $P$ and $f$ is controlled by the helicity density $h$. Furthermore, by integrating over the variable $s$, we can calculate the shape of the distribution $\mathcal{F}\left(x,y,z\right)$ in the initial coordinates $(x,y,z)$:

\begin{equation}\label{eq59}
\mathcal{F}=\int{fds}=\boldsymbol{w}\cdot\boldsymbol{D}\int{Pds}+h\int{Psds}.
\end{equation}

\noindent Let us now calculate the form of the distributions at thermodynamic equilibrium.
Since $vol_{\boldsymbol{y}}^{4}$ is the preserved volume element of a symplectic manifold spanned by canonical variables, we can exploit the usual formulation of statistical mechanics and define the differential entropy $\Sigma$ of the distribution function $P$: 

\begin{equation}\label{eq60}
\Sigma=-\int_{V_{\boldsymbol{y}}}{ P\log P\,vol_{\boldsymbol{y}}^{4}}.
\end{equation}

\noindent Here the integral is performed on the whole phase space $V_{\boldsymbol{y}}$.
The total number of particles and the total energy $E$ of the ensemble are given by $N=\int_{V_{\boldsymbol{y}}}{P\,vol_{\bol{y}}^{4}}$ and $E=\int_{V_{\bol{y}}}{HP\,vol_{\bol{y}}^{4}}$ respectively. 
The form of the distribution function at equilibrium is calculated my maximizing the entropy $\Sigma$ under the constraints $N$ and $E$ with the variational principle:

\begin{equation}\label{eq61}
\delta\left(\Sigma-\alpha N-\beta E\right)=0.
\end{equation}

\noindent Here $\alpha$ and $\beta$ are the Lagrange multipliers associated to $N$ and $E$. The result of the variation is:

\begin{equation}\label{eq62}
P=\frac{1}{Z}e^{-\beta H}.
\end{equation}

\noindent In the above equation $Z$ is a normalization constant. 
Thus, recalling equations (\ref{eq58}) and (\ref{eq59}), we
arrive at the following formulas for $f$ and $\mathcal{F}$:

\begin{subequations}\label{eq63}
\begin{align}
f&=\frac{1}{Z}\left(\boldsymbol{w}\cdot\boldsymbol{D}+sh\right)e^{-\beta H},\\
\mathcal{F}&=\frac{\Delta s}{Z}\left(\boldsymbol{w}\cdot\boldsymbol{D}+\frac{\Delta s}{2}h\right)e^{-\beta H}.
\end{align}
\end{subequations}

\noindent Here $\Delta s=\int{ds}$. The conclusion is that the thermodynamic equilibrium of a three dimensional ensemble governed by an antisymmetric operator departs from the standard Boltzmann distribution of homogeneous probability density on constant energy surfaces.
The distortion is controlled by the helicity density $h$, i.e. by the failure of the Jacobi identity. 

\subsection{\textit{Thermodynamic equilibrium in a non-integrable magnetic field}}

Consider an ensemble of magnetized particles
moving by $\bol{E}\cp\bol{B}$ drift according to equation (\ref{eq6}). 
The magnetic field $\boldsymbol{B}$ is assumed to be of the form:

\begin{equation}\label{eq64}
\boldsymbol{B}=\partial_{x}+\left(\frac{y-\sin y\cos y}{2}-\sin x\right)\partial_{z}.
\end{equation}  

\noindent One can verify that $\mathfrak{B}=d\left[\left(\frac{\sin^{2}y-y^{2}}{4}+y\sin x\right)dx+ydz\right]$.
Recalling that the antisymmetric Poisson operator is $\boldsymbol{w}=\boldsymbol{B}/B^{2}$, we have:

\begin{equation}\label{eq65}
\boldsymbol{w}=\frac{\partial_{x}+\left(\frac{y-\sin y\cos y}{2}-\sin x\right)\partial_{z}}{1+\left(\frac{y-\sin y\cos y}{2}-\sin x\right)^{2}},
\end{equation}

\noindent and also,

\begin{equation}\label{eq66}
h
=\frac{\sin^{2}y}{\left[1+\left(\frac{y-\sin y\cos y}{2}-\sin x\right)^{2}\right]^{2}}.
\end{equation}

\noindent A typical scenario encountered in magnetized plasmas is quasi-neutrality.
In such situation, the electric potential $\phi$ is, on average, zero. 
Therefore, the Hamiltonian of each `massless' particle is itself zero $H=\phi=0$. 
However, electrostatic fluctuations $\delta\phi$ generated by random interactions among charged particles drive the ensemble toward equilibrium, which according to (\ref{eq63}) is:

\begin{equation}\label{eq67}
\mathcal{F}=\frac{\Delta s}{Z}\left\{1+\frac{\Delta s}{2}\frac{\sin^{2}y}{\left[1+\left(\frac{y-\sin y\cos y}{2}-\sin x\right)^{2}\right]^{2}}\right\}.
\end{equation} 
 
\noindent Here we used equation (\ref{eq66}) and set $\boldsymbol{D}=\boldsymbol{B}=\boldsymbol{w}/w^{2}$ as required in the case of $\boldsymbol{E}\times\boldsymbol{B}$ drift. 
Figure \ref{fig3} shows a plot of the predicted thermal equilibrium.
The shape of the distribution sensibly departs from the flat profile 
one would expect by a naive application of the entropy principle in the initial (non-Hamiltonian) coordinates.
This discrepancy is a consequence of the failure of the Jacobi identity.  


\begin{figure}[h]
\hspace*{0cm}\centering
\includegraphics[scale=0.14]{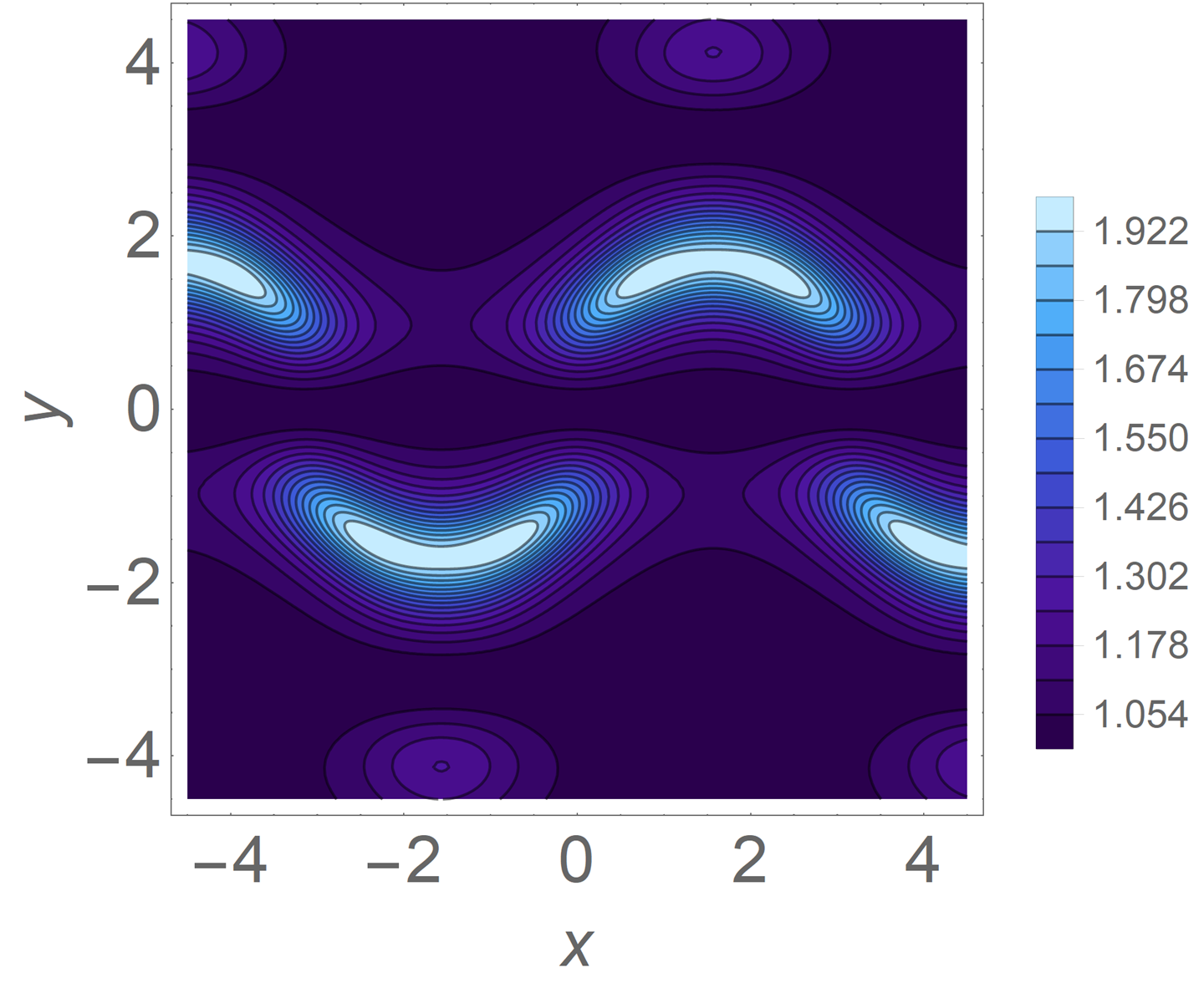}
\caption{\footnotesize Thermal equilibrium $\mathcal{F}\left(x,y\right)$ by $\boldsymbol{E}\times\boldsymbol{B}$ drift in the magnetic field (\ref{eq64}). The inhomogeneous distribution is caused by the failure of the Jacobi identity.}
\label{fig3}
\end{figure}	

\section{Concluding Remarks}
\label{sec:6}

Antisymmetric operators arise in the presence of non-integrable constraints. 
In this study we investigated the problem of repairing an
antisymmetric operator in order to recover a Poisson structure.

This issue is encountered when formulating the statistical mechanics of 
conservative ensembles. Indeed, due to the
absence of the invariant measure guaranteed by Liouville's theorem in the Hamiltonian framework, 
the usual arguments of statistical mechanics cannot be applied. 

In order to overcome these problems, we devised a systematic method to
adjust a three dimensional antisymmetric operator by extending it to four dimensions
and by operating a proper time reparametrization. 
The Poissonization procedure does not rely on the existence of special
symmetries and therefore does not rely on the specific form of the Hamiltonian
function.

A concrete example pertaining to the motion of a charged particle in a
magnetic field has been examined. By Poissonizing the system we were able
to construct a canonical phase space where the motion is regular and incom-
pressible. A physical interpretation of the extended variable and of the proper
time was also given: they mimic the motion along the magnetic
field.

Finally, we determined the Jacobian of the coordinate change linking the initial
coordinates to the extended canonical phase space. We found that the Jacobian is controlled
by the helicity density of the antisymmetric operator, which measures the failure of the Jacobi identity. 
By invoking the principle of  maximum entropy 
in the novel phase space we constructed the distribution function of
thermodynamic equilibrium: in the a priori coordinates the shape of the distribution departs
from the standard Boltzmann distribution due to the finite helicity density of the
antisymmetric operator. The theory has been applied to obtain the equilibrium distribution of an ensemble of charged particles moving by $\bol{E}\cp\bol{B}$ drift.

\section{Acknowledgments}
\label{sec:ack}
The author would like to acknowledge useful discussion with Professor Z. Yoshida.
The work of N. S. was supported by JSPS KAKENHI Grant No. 16J01486 and No. 18J01729.




\end{normalsize}

\end{document}